\documentclass[%
preprint,
preprintnumbers,
 amsmath,amssymb,
 aps,
]{revtex4-2}

\usepackage{graphicx}
\usepackage{bm}
\usepackage[ruled,vlined]{algorithm2e}
\usepackage{hyperref}

\usepackage{xcolor}

\newcommand{\adjointsource}{s^\dagger}

\newcommand{\linearforward}{\mathcal{N}_{\boldsymbol{q}}}
\newcommand{\sourceshort}{S}
\newcommand{\measkernel}{\mathcal{\zeta}}

\newcommand{\source}{\sourceshort(\boldsymbol{x};\boldsymbol{x}_s)}

\begin{document}

\preprint{APS/123-QED}

\title{Localization of sources in weakly nonlinear fluid systems using linear and quadratic sensitivity analysis}

\author{Qi Wang}%
 \email{qwang4@sdsu.edu}
\author{Zejian You}%
\affiliation{%
 Department of Aerospace Engineering, San Diego State University, San Diego, California 92182, USA}%
\begin{abstract}
We develop a framework for localized source detection in dynamical systems governed by nonlinear partial differential equations based on first and second-order sensitivity analysis. Building on the standard adjoint formulation, which relates multiple measurements to external sources through a linear duality relation, we first introduce a linear positional embedding that identifies the source location by aligning the measurement vector with the embedding. To capture weakly nonlinear effects that arise when the source intensity is finite, we then incorporate a quadratic correction represented as a symmetric bilinear operator and approximated via a truncated eigen-expansion obtained with Krylov subspace iterations. This yields quadratic positional embeddings that augment the linear adjoint field, enabling measurement data to be projected onto a higher-dimensional hyperplane, spanned by the linear and quadratic embeddings. A source search algorithm is formulated based on principal angle minimization between this hyperplane and the observation vector, providing a natural probabilistic interpretation of source location. The method operates in a one-shot fashion without iterative updates of candidate source positions, and it can be readily extended to scenarios involving multiple sources. Demonstrations on benchmark inverse problems include perturbation-source identification in the viscous Burgers equation and heat-source detection in a two-dimensional laminar stratified channel. The results with quadratic embeddings show significant improvements in localization accuracy compared with linear adjoint-based sensitivity methods, especially in the region where linear adjoint sensitivity vanishes.
\end{abstract}

\maketitle

\section{Introduction}
Sensitivity analysis is a cornerstone of solving inverse problems, particularly in fields such as fluid dynamics, heat transfer, and control systems \citep{zaki2024turbulence,wang2013drag,hall1983physical, dominguez2025inference}. Traditional approaches often rely on linearized tangent models, which assume small perturbations and near-linear system behavior. A central concept in these methods is the domain of dependence \citep{wang2025domain, wang_wang_zaki_2022}, typically computed via the adjoint operator, which identifies \emph{where} and \emph{when} perturbations in a system would influence a given measurement. This domain plays a critical role in establishing causal relationships and in interpreting observational data.
While linear sensitivity analysis is effective in many settings, it becomes inadequate for systems exhibiting nonlinear dynamics or when perturbations are of finite amplitude. 

In the current work, we develop a positional embedding framework for source detection that generalizes classical sensitivity analysis. A linear positional embedding is first defined from the adjoint field, which enables one-shot source reconstruction using the alignment of measurement vectors with candidate source locations. To capture weakly nonlinear effects, we augment this with a quadratic positional embedding, constructed via a low-rank approximation of a bilinear correction operator obtained through an offline Krylov subspace iteration. Together, these embeddings provide a compact representation that allows source localization without iterative updates of candidate positions during inference. The framework is examined on two canonical problems, and is further tested on the simultaneous reconstruction of two heat sources in temperature-stratified channel flow. These demonstrations highlight the ability of quadratic sensitivity analysis to recover source locations more accurately than linear adjoint-based methods in weakly nonlinear regimes.

\subsection{source identification method}
Identifying localized sources in fluid systems is complicated by the irreversible loss of information inherent to many physical processes \citep{wang2019spatial,tanogami2024scale,antenucci2001energetics}. 
Scalar plumes in flow fields are distorted by diffusion and dispersion, making inverse reconstruction ill-conditioned. 
In temperature-stratified flows, thermal sources can excite internal gravity waves and lee waves that propagate over long distances, but their extended reach complicates the attribution of measurements to specific sources \citep{sutherland2010internal,staquet2002internal}. 
In wall-bounded flows, which are common in both geophysical and engineering settings, 
boundary layers act as low-pass filters that suppress small-scale fluctuations, further obscuring the fine details needed for accurate source localization \citep{hasegawa2009low}.

Early approaches to source reconstruction relied on forward, trial-and-error strategies in which candidate sources were postulated and simulated, with their outputs compared against measurements \citep{liu2007inverse}. 
For simple or steady flows, this approach allowed the explicit formation of an impulse-response matrix mapping source inputs to sensor outputs \citep{gorelick1983identifying, alapati2000recovering, WRCR:WRCR6355, mahar1997optimal, mahar2000identification, mahar2001optimal}. 
However, the high dimensionality and complexity of realistic turbulent flows render such brute-force methods intractable \citep{misaka2008measurement}.  
Adjoint-based methods offered a breakthrough by enabling the efficient computation of source sensitivities without explicitly constructing the impulse-response operator. 
Within this adjoint-variational framework, the gradient of a data--model mismatch cost function with respect to the source distribution can be obtained through a single forward and adjoint solve. 
This approach has been widely adopted in contexts ranging from numerical weather prediction and tsunami inversion to flow reconstruction in wall-bounded turbulence \citep{olaguer2011adjoint,courtier1997use, pires2003sensitivity, li2013optimization, kim2017effect, cerizza2016reconstruction, wang_hasegawa_zaki_2019, abassi2025adjoint, mengze2019discrete}.  
When observations are noisy or incomplete, probabilistic methods provide an alternative pathway. 
Bayesian inference frameworks, for example, have been applied to diverse problems ranging from pollution detection in groundwater systems \citep{keats2007bayesian} to vortex identification from sparse pressure signals, where they 
yield plausible flow structures even in underdetermined scenarios \citep{eldredge2024bayesian}.

Despite these advances, traditional physics-based inference approaches face two central challenges. First, they require repeated full-physics simulations during inference, which is computationally prohibitive in practice. Second, they struggle with determining and extracting the “right amount” of information: sufficient to enable robust source reconstruction while avoiding overfitting or spurious solutions. These difficulties are not limited to turbulent flows but arise broadly in nonlinear systems where diffusion, dispersion, or incomplete background knowledge leads to an irreversible loss of recoverable information. Such challenges motivate the development of alternative frameworks that can efficiently encode source signatures in a low-dimensional representation and enable accurate inference without iterative forward–inverse cycles.

\subsection{Linear and quadratic sensitivity analysis using adjoint}
A key strength of the adjoint formulation is that it reveals the domain of dependence of a measurement—the region in space and time that causally influences the observable. This information not only provides crucial information for applications such as source detection, data assimilation, sensor placement, and flow control \citep{wang2019spatial, wang_wang_zaki_2022, wang2025domain}, but also delineates the limits of inference in more complex flows \citep{zaki2025arfm,zaki2021prfaps,mengze2021}.

Classical (first-order) adjoint methods relate small source perturbations to measurement variations through a linear duality, enabling gradient evaluation at the cost of one forward–adjoint pair of solves. When source amplitudes are finite or weakly nonlinear effects matter, linear sensitivity underestimates the true response, and a \emph{quadratic} correction becomes necessary. Second-order adjoint (SOA) methodology provides this correction by yielding Hessian–vector products or, in continuous form, a symmetric bilinear map that captures second-order variations of the data–model misfit or of sensor functionals with respect to inputs \citep{wang1992second, cioaca2012second, le2021second, cacuci2016second}.
SOA constructs the Hessian-vector product for arbitrary directions at a cost comparable to a small multiple of a forward–adjoint pair, avoiding explicit Hessian assembly \citep{cioaca2012second,le2002second}. In hydrodynamic stability and flow control, related second‑order operators quantify quadratic eigenvalue sensitivity to steady controls, enabling maps of second-order effects without recomputing controlled base flows \citep{boujo2021second}.

The remainder of the paper is organized as follows. In \S\ref{sec:embedding}, we introduce the linear and quadratic positional embeddings, which map each potential source location to a high-dimensional vector representation. In \S\ref{sec:identification}, we describe the algorithm used to identify source locations based on these embeddings. Sections \S\ref{sec:burgers} and \S\ref{sec:stratifiedchannel} present results for two benchmark problems: perturbation-source detection in the one-dimensional viscous Burgers equation, serving as a conceptual demonstration, and heat-source identification in a two-dimensional temperature-stratified laminar channel flow. Finally, in \S\ref{sec:discussion}, we extend the framework to the simultaneous reconstruction of two sources in the stratified channel flow.

\section{Methodology}
\label{sec:methodology}
Let $\mathbb{V}$ denote the Hilbert space of square-integrable scalar fields on the spatial domain $\Omega$. 
We write $\mathbb{V}^d$ for the $d$-fold Cartesian product space corresponding to $d$-component perturbation fields, i.e. $\boldsymbol{q} \in \mathbb{V}^d$. 
Here, $d$ denotes the number of state variables (e.g. velocity components and temperature), not the spatial dimension of $\Omega$.
The governing dynamics are modeled as a partial differential equation (PDE) with external forcing,
\begin{equation}
\mathcal{N}(\boldsymbol{q}(\boldsymbol{x},t)) 
= \mathbf{P}\, \source\, J(t),
\label{eq:governing_general}
\end{equation}
where $\mathcal{N}$ denotes the PDE operator, $\boldsymbol{q} \in \mathbb{V}^d \times [0,T]$ represents the $d$-component perturbation state variables, 
$\sourceshort(\boldsymbol{x};\boldsymbol{x}_s)\in \mathbb{V}$ is the spatial distribution of source,  
$J(t)\in [0, T]$ is the temporal profile of the source, and $\boldsymbol{x}_s \in \mathbb{R}^n$ is the spatial location of the external source to be identified. 
The operator $\mathbf{P}: \mathbb{V} \to \mathbb{V}^d$ is a constant \emph{lifting operator} that injects the scalar source term into the appropriate component of $\boldsymbol{q}$; its transpose, $\mathbf{P}^{\top}: \mathbb{V}^d \to \mathbb{V}$, acts as a \emph{projection} extracting the corresponding component from the state. 
For instance, in a temperature-stratified flow, one may take $\mathbf{P}=(0,0,1)^{\top}$ if the third component of $\boldsymbol{q}$ corresponds to temperature. 
The temporal profile $J(t)$ is assumed known---either impulsive (modeled as a Dirac $\delta$-function in time) or steady (modeled as a Heaviside function switched on at the initial time). 
For spatial localization, we adopt sources with Gaussian profiles of finite width rather than idealized delta functions. 
The particular source shape does not affect the accuracy of the proposed algorithm; the Gaussian form is chosen for analytical and numerical convenience. 
In this formulation, the operator $\mathcal{N}$ incorporates both the dynamics and initial conditions in a lifted form.
More importantly, for notation convenience, the operator $\mathcal{N}$ is \emph{homogenized} with respect to the source {$\sourceshort$}, meaning that it represents the nonlinear perturbation equation, where the baseline solution corresponding to {$\sourceshort=0$} has been subtracted.

\subsection{First and second-order sensitivity analysis and positional embeddings}
\label{sec:embedding}
We now develop the concept of \emph{positional embeddings}, which map each candidate source location to a high-dimensional vector that can be directly compared with measurement data. These embeddings are constructed by exploiting adjoint sensitivity analysis.  
For arbitrary variables $\boldsymbol{v}_1, \boldsymbol{v}_2 \in \mathbb{V}^d \times [0,T]$, we define the spatiotemporal inner product as,
\[
\left\langle \boldsymbol{v}_1(\boldsymbol{x},t), \, \boldsymbol{v}_2(\boldsymbol{x},t) \right\rangle_{\mathbb{V}^d,t}
= \int_0^T \int_{\Omega} \boldsymbol{v}_1^{\top}(\boldsymbol{x},t)\,\boldsymbol{v}_2(\boldsymbol{x},t)\, d\boldsymbol{x}\, dt .
\]
In particular, for scalar fields $f_1,f_2 \in \mathbb{V}$ we define the spatial inner product
\[
\left\langle f_1(\boldsymbol{x}), f_2(\boldsymbol{x}) \right\rangle
= \int_{\Omega} f_1(\boldsymbol{x}) f_2(\boldsymbol{x}) \, d\boldsymbol{x}.
\]

Let $\mathbb{R}^{N_m}$ denote the measurement space, where $N_m$ is the number of independent sensors or observation channels.
Measurements are defined as spatiotemporal integrations over the state variables $\boldsymbol{q}$ using a collection of measurement kernels
$\boldsymbol{M} = (M_1,\dots,M_{N_m})$,
\begin{equation}
\boldsymbol{m}
=
\langle \boldsymbol{q}, \boldsymbol{M} \rangle_{\mathbb{V}^d \times t}.
\end{equation}
This spatiotemporal integration framework encompasses a wide class of measurement operators. For example, pointwise measurements correspond to choosing $M_j$ as products of spatial and temporal Dirac delta functions, in which case the inner product reduces to evaluating $\boldsymbol{q}$ at a prescribed location and time. More general measurements are obtained by selecting smoother kernels that integrate $\boldsymbol{q}$ over finite spatial regions,
time intervals, or both.
For infinitesimal perturbations, higher-order terms may be neglected and the forward operator $\mathcal{N}$ is identified as the
\emph{linearized forward operator}, denoted by $\linearforward$.
This operator $\linearforward$ is precisely the Fréchet derivative \citep{zeidler2013nonlinear} of the forward equation evaluated at the baseline state and its inverse $\linearforward^{-1}$ corresponds to the solution operator
of the linearized forward equations, namely
\begin{equation}
\boldsymbol{q}(\boldsymbol{x},t)
=
\linearforward^{-1}\!\left(
\mathbf{P}\sourceshort\, J(t)
\right),
\end{equation}
where $\mathbf{P}$ denotes the source-to-state mapping and $J(t)$ is a prescribed temporal modulation.
Substituting this expression into the spatiotemporal integration yields
\begin{equation}
\label{eqn:linear_sensitivity_new}
\begin{aligned}
\boldsymbol{m}
&=
\left\langle
\linearforward^{-1}
\big(
\mathbf{P}\sourceshort J(t)
\big),
\boldsymbol{M}
\right\rangle_{\mathbb{V}^d \times t} \\
&=
\left\langle
\mathbf{P}\sourceshort J(t),
\boldsymbol{q}^\dagger(\boldsymbol{x},t)
\right\rangle_{\mathbb{V}^d \times t}
=
\left\langle
\sourceshort,
\boldsymbol{\adjointsource}
\right\rangle_{\mathbb{V}}=\boldsymbol{s}^\dagger[\sourceshort],
\end{aligned}
\end{equation}
where $\boldsymbol{q}^\dagger = (\boldsymbol{q}_1^\dagger,\ldots,\boldsymbol{q}_{N_m}^\dagger)$ denotes the adjoint states. These adjoint states satisfy
the adjoint equations
\begin{equation}
\linearforward^\dagger
\big(
\boldsymbol{q}_j^\dagger
\big)
=
M_j,
\qquad
j = 1,\ldots,N_m,
\end{equation}
with $\linearforward^\dagger$ denoting the adjoint of the linearized forward operator.
The adjoint source $\boldsymbol{\adjointsource}$ is given componentwise by
\begin{equation}
\adjointsource_j(\boldsymbol{x})
=
\int_0^T
J(t)\,
\mathbf{P}^\top
\boldsymbol{q}_j^\dagger(\boldsymbol{x},t)
\, dt,
\qquad
j = 1,\ldots,N_m,
\end{equation}
which corresponds to time-weighted projections of the adjoint states onto the source space.
From a functional-analytic perspective, the measurement operator
$\boldsymbol{\measkernel} : \mathbb{V} \to \mathbb{R}^{N_m}$ maps the source $\sourceshort \in \mathbb{V}$ to the measurement vector
$\boldsymbol{m} \in \mathbb{R}^{N_m}$. When the perturbation of the source is infinitesimal, $\boldsymbol{\measkernel}$ is linear and belongs to the dual space $(\mathbb{V})^\dagger$.
The adjoint variable $\boldsymbol{s}^\dagger$ is the unique Riesz representer of this functional $\boldsymbol{\measkernel}$ in the Hilbert space $\mathbb{V}$ \citep{conway2019course}.
This Riesz representer defines a \emph{linear positional embedding}: for a fixed source shape, each candidate source location is mapped to a
high-dimensional vector. Source identification may therefore be interpreted as aligning the measurement vector $\boldsymbol{m}$ with these embeddings.

When the perturbation amplitude is finite, however, the purely linear duality
\eqref{eqn:linear_sensitivity_new} no longer holds. Weak nonlinear interactions introduce systematic deviations that must be accounted for to recover
accurate source information. To capture these effects, we introduce a \emph{quadratic correction} to the duality relation, analogous to a second-order
Taylor expansion,
\begin{equation}
\label{eqn:quadratic_embedding}
\boldsymbol{m}
=
\boldsymbol{\measkernel}(\sourceshort)
\;\approx\;
\boldsymbol{s}^\dagger[\sourceshort]
\;+\;
\frac{1}{2}\,
\mathcal{H}[\sourceshort,\sourceshort],
\end{equation}
where $\mathcal{H} : \mathbb{V} \times \mathbb{V} \to \mathbb{R}^{N_m}$ is the symmetric bilinear form associated with the second Fréchet derivative
(the Hessian) of the measurement operator. This correction defines a \emph{quadratic positional embedding}, which augments the linear embedding and captures
the leading-order nonlinear effects.

In practice, $\mathcal{H}$ is too high-dimensional to assemble explicitly. 
We therefore approximate it using a low-rank eigen-expansion,
\begin{equation}
    \mathcal{H}_j \;\approx\; \sum_{k=1}^{N_{eig}} \lambda_k \, \psi_{j,k}(\boldsymbol{x}) \, \psi_{j,k}(\boldsymbol{x}^{\prime}), 
    \qquad \langle \psi_{j,k_1}, \psi_{j,k_2} \rangle = \delta_{k_1, k_2}, \quad 1 \le k_1, k_2 \le N_{eig}, \quad 1 \le j \le N_m,
\end{equation}  
where $\{ \psi_{j,k} \}$ are orthonormal eigenmodes of $\mathcal{H}_j$, the Hessian for the $j$-th measurement, and $\{ \lambda_{j,k} \}$ are the corresponding eigenvalues. 
The dominant eigenmodes are computed efficiently by a Krylov subspace iteration procedure, which extracts the leading contributions to $\mathcal{H}$ without requiring its full construction.

We briefly state the approximation of Hessian-source multiplication $H_j(v)$ for the $j$-th measurement kernel.
Recall that the linear Riesz representer for the $j$-th measurement is $s^\dagger_j := \int_0^T J(t)\,\mathbf{P}^{\top}\linearforward^{-\dagger}\big({M}_j\big)\,dt$.
To obtain the quadratic (Hessian) correction in the direction $v\in\mathbb{V}$, we perturb the baseline source $S=0$ by a small amount $\varsigma v$,
lift it with $\mathbf{P}$, and compute the corresponding nonlinear perturbed state from $\sourceshort=0 + \varsigma v$,
$\boldsymbol{q}_\epsilon := \mathcal{N}^{-1}(\mathbf{P}\,\varsigma v J(t))$.
The change in the Riesz representer induced by this perturbation is
\begin{equation}
    \Delta s^\dagger_j
    = \int_0^T J(t)\,\mathbf{P}^{\top}\!\mathcal{N}^{-\dagger}_{\boldsymbol{q}_\epsilon}\big({M}_j\big)\,dt - s_j^{\dagger},
\end{equation}
which, upon division by $\varsigma$ and taking $\varsigma\to 0$, yields the directional derivative of $\boldsymbol{s}^\dagger_j$ along $v$, i.e. the Hessian–vector action. We therefore define
\begin{equation}
    H_j(v)
    := \frac{1}{\varsigma}\left(\Delta s^{\dagger}_j\right)
    = \frac{1}{\varsigma}\!\left(\int_0^T J(t)\,\mathbf{P}^{\top}\,\mathcal{N}^{-\dagger}_{\boldsymbol{q}_\epsilon}\big({M}_j\big)\,dt \;-\; s^\dagger_j\right).
    \label{eqn:Hv_new}
\end{equation}
This matrix-free formula realizes the second Fr\'echet derivative of the measurement map in the direction $v$ without assembling the Hessian: it requires one perturbed forward solve to obtain $\boldsymbol{q}_\epsilon$ and one adjoint solve associated with the operator linearized at $\boldsymbol{q}_\epsilon$ to evaluate $\mathcal{N}^{-\dagger}_{\boldsymbol{q}_\epsilon}\big({M}_j\big)$.
With this approximation of Hessian-vector multiplication, we can adopt standard Krylov subspace iteration methods to compute the eigenpairs of $\mathcal{H}$ without explicitly formulating the operator \citep{parlett1998symmetric,lehoucq1998arpack}.
The overall procedure for constructing the quadratic sensitivity modes is summarized in algorithm~\ref{alg:subspace-noisy}. 
In the present study, 
The Hessian--vector product in \eqref{eqn:Hv_new} is approximated using finite differences with a convergence tolerance of $\epsilon = 10^{-3}$ and a perturbation amplitude of $\varsigma = 10^{-4}$. 
This choice is motivated primarily by computational efficiency considerations. 
Assume that the forward and adjoint simulations have comparable computational cost.
Under this assumption, a finite-difference approximation of the Hessian--vector product---which requires one perturbed forward solve and two adjoint solves---has a computational cost of the same order as a single adjoint simulation per Krylov iteration. 
Since the number of Krylov iterations required for convergence is typically much smaller than the number of degrees of freedom in the discretized domain, the overall cost remains scalable for large-scale problems.
If a differentiable solver is available, the Hessian--vector product can be achieved by the forward-over-reverse automatic differentiation strategy.
This product, as shown here, is approximated by finite difference over the adjoint gradients.
For large-scale PDE systems, an exact automatic-differentiation implementation of Hessian--vector products requires propagating through the forward time integration in addition to a backpropagation, which can significantly increase memory usage and computational cost. 
The finite-difference-over-adjoint approach only requires repeated gradient evaluations and is therefore more practical for the present study.

\begin{algorithm}
\caption{Subspace iteration with symmetric matrix--vector products}
\label{alg:subspace-noisy}
\KwIn{Target eigenpairs $N_{eig}$, Krylov size $Q>N_{eig}$, max restarts $R_{\max}$, residual tolerance $\varepsilon$, breakdown tolerance $\tau_{\mathrm{break}}$, measurement index $j$}
\KwOut{Approximate leading eigenpairs $\{(\lambda_{j,k},\psi_{j,k})\}_{k=1}^{N_{eig}}$}

Choose a random unit--norm start $u_1$; locked set $\mathcal L \leftarrow \emptyset$.\;
\For{$r=1$ \KwTo $R_{\max}$}{
  \If{$\mathcal L\neq\emptyset$}{ orthogonalize $u_1$ against locked vectors and renormalize }
  \For{$i=1$ \KwTo $Q$}{
    $w \leftarrow H_j(u_i)$ \tcp*{approximate Hessian-vector multiplication}
    \For{$p=1$ \KwTo $i$}{ $w \leftarrow w - \langle u_p,w\rangle u_p$ }
    $\beta \leftarrow \|w\|$\;
    \If{$\beta \le \tau_{\mathrm{break}}$}{ set $Q_{\mathrm{eff}} \leftarrow i$; \textbf{break} }
    $u_{i+1} \leftarrow w/\beta$, $Q_{\mathrm{eff}}\leftarrow i+1$\;
  }
  $\mathbf{U} \leftarrow [u_1,\ldots,u_{Q_{\mathrm{eff}}}]$\;
  $\mathbf{G} \leftarrow \tfrac{1}{2}\big(\mathbf{U}^\top H_j(\mathbf{U})+[\mathbf{U}^\top H_j(\mathbf{U})]^\top\big)$ \tcp*{symmetrize Gram matrix}
  Compute $\mathbf{GY}=\mathbf{Y\Lambda}$ and sort eigenpairs by decreasing $|\lambda_{j,k}|$\;
  $\mathbf{\Psi} \leftarrow \mathbf{UY}$, $\lambda_k \leftarrow \Lambda_{kk}$\;
  \For{$k=1$ \KwTo $\min(N_{eig},Q_{\mathrm{eff}})$}{
    \If{$\|H_j(\psi_{j,k})-\lambda_{j,k}\psi_{j,k}\|<\varepsilon$}{ add $(\lambda_{j,k},\psi_{j,k})$ to $\mathcal L$ }
  }
  \If{$|\mathcal L|\ge N_{eig}$}{ \textbf{break} }
  Select $p$ indices among unconverged modes with largest $|\lambda_{j,k}|$; form $\mathbf{U}_{\mathrm{new}}$.\;
  Orthonormalize $\mathbf{U}_{\mathrm{new}}$ and reorthogonalize against $\mathcal L$.\;
}
\Return the best $N_{eig}$ pairs from $\mathcal L$.
\end{algorithm}

\subsection{Source search using linear and quadratic positional embeddings}
\label{sec:identification}

Analogous to a second-order Taylor expansion, the quadratic positional embedding in Equation~\eqref{eqn:quadratic_embedding} provides an approximation of the measurement response with respect to the source $\sourceshort$.
Given the large storage required for the Hessian, it is preferable to approximate it in terms of the leading eigenpairs 
$\{(\lambda_k,\psi_{j,k})\}_{k=1}^{N_{\mathrm{eig}}}$. With such approximation, equation \eqref{eqn:quadratic_embedding} can be rewritten as
\begin{equation}
\begin{aligned}
m_j 
&\approx 
I_s \underbrace{\left\langle K(\boldsymbol{x}_s), s_j^{\dagger} \right\rangle}_{\tilde{s}_j^{\dagger}(\boldsymbol{x}_s)}
+ \frac{1}{2} I_s^2 
\sum_{k=1}^{N_{\mathrm{eig}}} \lambda_k 
\Big[
\underbrace{
\left\langle \psi_{j,k}(\boldsymbol{x}_s), K(\boldsymbol{x}_s) \right\rangle
}_{\tilde{\psi}_{j,k}(\boldsymbol{x}_s)}
\Big]^2 \\
&= I_s\, \tilde{s}_j^{\dagger}(\boldsymbol{x}_s)
+ \frac{1}{2} I_s^2 
\underbrace{
\sum_{k=1}^{N_{\mathrm{eig}}} \lambda_k 
\tilde{\psi}_{j,k}^2(\boldsymbol{x}_s)
}_{\tilde{h}_j(\boldsymbol{x}_s)} ,
\end{aligned}
\label{eq:taylor_expansion}
\end{equation}
where the source is expressed as the product of the unknown source
intensity $I_s$ and the source shape 
$K(\boldsymbol{x};\boldsymbol{x}_s)$ centered at $\boldsymbol{x}_s$.
For notational convenience, we abbreviate the source shape function,
\begin{equation}
\source = I_s\, K(\boldsymbol{x}_s).
\end{equation}

\begin{figure}
    \includegraphics[width=0.9\linewidth]{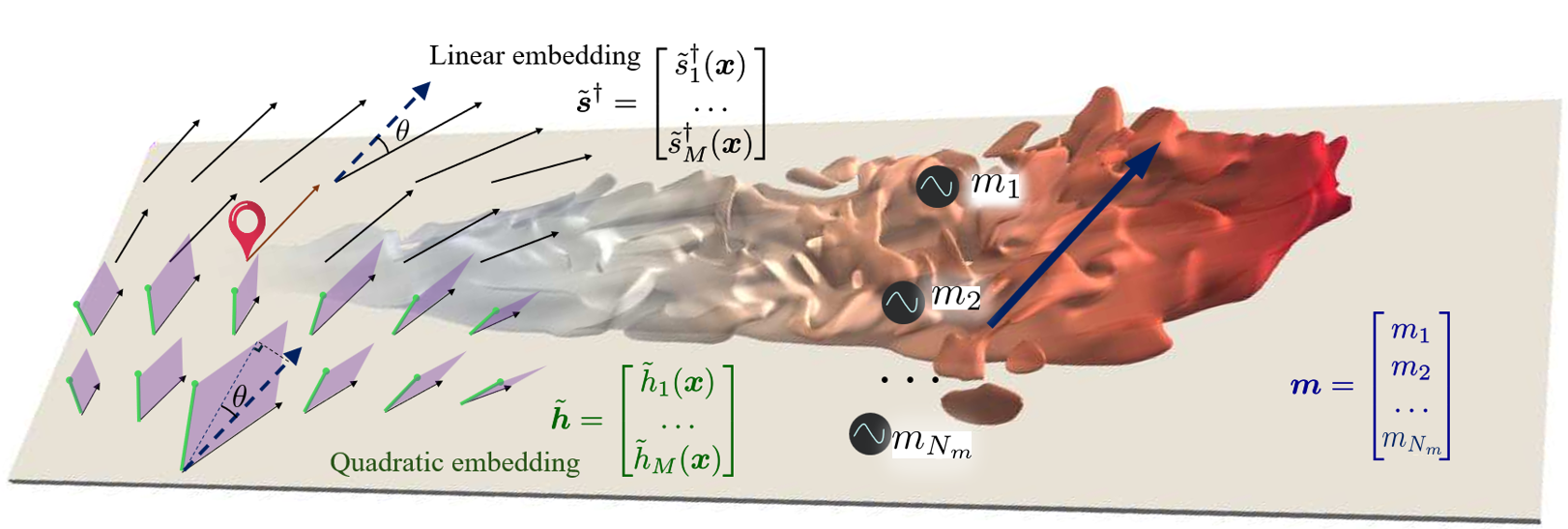}
    \vspace{-20 pt}
    \caption{Schematic of the source-localization framework using positional embeddings. 
    A set of downstream sensors ($m_1, m_2, \ldots, m_{N_m}$) records signals whose concatenation forms the measurement vector $\boldsymbol m$, shown as a blue vector.
    For each candidate upstream source location $\boldsymbol{x}_s$, the adjoint fields from the measurement kernels define a linear positional embedding $\tilde{\boldsymbol s}^\dagger(\boldsymbol{x}_s)$ (black), while the approximations of the Hessian define the quadratic positional embeddings $\tilde{\boldsymbol h}(\boldsymbol{x}_s)$ (green). Source location is identified where the measurement vector $\boldsymbol{m}$ falls in the linear space spanned by the linear and quadratic embeddings. The principal angle $\theta$ between $\boldsymbol m$ and its projection quantifies the consistency of the candidate source location with the observed measurements.}
    \label{fig:schematic}
\end{figure}

For a purely localized source, i.e., $K(\boldsymbol{x}_s) = \delta(\boldsymbol{x} -\boldsymbol{x}_s)$, the first and second inner product on the right-hand side would be $s_j^{\dagger}(\boldsymbol{x}_s)$ and $\psi_{j,k}(\boldsymbol{x}_s)$. When the shape of the source is a more regular shape (say, a Gaussian), we denote the results as the tilde quantities.
When the source intensity is infinitesimal, the quadratic correction is negligible and source localization reduces to assessing the alignment between the measurement vector $\boldsymbol{m}$ and the linear embedding direction. This regime is illustrated in the upper-left panel of
Fig.~\ref{fig:schematic}, where a consistent source location corresponds to the two vectors being parallel.
When the source intensity is finite, nonlinear effects become significant and the measurement generally deviates from the linear prediction.
In this case, the source is identified as the location for which the measurement vector lies within the span of
$\tilde{\boldsymbol{s}}^{\dagger}$ and $\tilde{\boldsymbol{h}}$, as illustrated in the lower-left panel of the schematic.
The exact relationship between coefficients for $\tilde{\boldsymbol{s}}^\dagger$ and $\tilde{\boldsymbol{h}}$ is not enforced here, in order to absorb moderate model error in the exact quadratic form, especially when the embedding $\tilde{\boldsymbol{s}}^{\dagger}$ and $\tilde{\boldsymbol{h}}$ are nearly parallel. It reduces a nonconvex 1D curve‑fitting into a convex projection, which worked much better than using the exact form of quadratic fitting.
We define the orthogonal projector onto the span of the linear and quadratic
embeddings $B(\boldsymbol{x}_s)=[\,\tilde{\boldsymbol{s}}^{\dagger}(\boldsymbol{x}_s)\ \ \tilde{\boldsymbol{h}}(\boldsymbol{x}_s)\,]$ by
\[
\mathcal{P}(\boldsymbol{x}_s)
= B(\boldsymbol{x}_s)\!\left(B(\boldsymbol{x}_s)^\top B(\boldsymbol{x}_s)\right)^{-1}\! B(\boldsymbol{x}_s)^\top.
\]
The projected vector $\mathcal P(\boldsymbol{x}_s)\boldsymbol m$, shown as a blue dashed vector in \ref{fig:schematic}, is the closest approximation to the measurement within this subspace.
The principal angle $\theta$ between $\boldsymbol m$ and its projection, together with the corresponding projection coefficients, is used to quantify the consistency of the candidate source location with the observed measurements,
\begin{equation}
\theta(\boldsymbol{x}_s)
:= \arccos\!\left(
\frac{\big\|\,\mathcal{P}(\boldsymbol{x}_s)\,\bm{m}\,\big\|}
     {\big\|\,\bm{m}\,\big\|}
\right)\in[0,\tfrac{\pi}{2}], \quad \boldsymbol{z} =\left(B(\boldsymbol{x}_s)^\top B(\boldsymbol{x}_s)\right)^{-1}\! B(\boldsymbol{x}_s)^\top \boldsymbol{m}.
\label{eqn:quad_angle}
\end{equation}
When only linear embedding is used,
\begin{equation}
\theta(\boldsymbol{x}_s)
:= \arccos\!\left(
\frac{\,\big|\boldsymbol{m}^{\top}\tilde{\boldsymbol{s}}^{\dagger}(\boldsymbol{x}_s)\big|}
     {\|\tilde{\boldsymbol{s}}^{\dagger}(\boldsymbol{x}_s)\|\,\|\bm{m}\|}
\right).
\label{eqn:linear_angle}
\end{equation}
The probability of source location $\boldsymbol{x}_s$ can be naturally defined as,
\begin{equation}
    P(\boldsymbol{x}_s) \propto \exp{(-\gamma \theta(\boldsymbol{x}_s))} P(\boldsymbol{z}),
    \label{eqn:probability}
\end{equation}
where $\theta(\boldsymbol{x}_s)\in[0, \pi/2]$ denotes the angle between the measurement vector and the subspace spanned by the linear and quadratic embeddings, and 
$P(\boldsymbol{z})$ represents the prior distribution of the projection coefficients. For example, the source intensity $z_1$ and the quadratic coefficient $z_2$ are restricted to be non-negative, motivating a prior of the form $P(\boldsymbol{z}) = \prod_i e^{-z_i}\,\mathbf{1}_{{z_i>0}}$, i.e., exponential distributions with unit rate.
The exponential prior distribution of the coefficient vector $\boldsymbol{z}$ is constructed to impose a small-magnitude bias under the weakly nonlinear assumption. Given that the source intensity is finite, a quadratic positional embedding provides an effective approximation of the resulting weakly nonlinear process.

The choice of an exponential likelihood in the angular deviation is primarily a phenomenological model rather than one derived from a specific uncertainty distribution (which belongs to future work). In principle, Gaussian measurement noise in Euclidean space would lead to angular statistics more closely related to von Mises–Fisher or Gaussian-in-angle distributions. Nevertheless, in the present study, we found that the exponential form provided excellent empirical performance across a range of test cases, with results that were more stable and sharply localized compared to alternative choices. For this reason, we adopt the exponential form here and defer a systematic investigation of the connection between noise models and angular likelihoods to future work.
We further define the Maximum A Posteriori (MAP) estimation \citep{bishop2006pattern},
\begin{equation}
    \boldsymbol{x}_s^{\mathrm{MAP}}
    = \operatorname*{arg\,max}_{\boldsymbol{x}_s}
    \; \exp\bigl(-\gamma \theta(\boldsymbol{x}_s)\bigr)\,
      P(\boldsymbol{z}) .
\end{equation}
In the present study, we set the hyperparameter $\gamma$ to $20$. 
This choice implies that a measurement vector deviating by $10^\circ$ from the hyperplane spanned by the linear and quadratic embeddings receives only about $3\%$ of the weight of a perfectly aligned vector. 
Thus, $\gamma$ controls the angular tolerance of the inference procedure, with larger values concentrating probability more tightly around the embedding subspace.

\section{Results}
\label{sec:results}
We report results from two canonical configurations designed to assess the performance and generality of the proposed framework. The first configuration involves a localized external impulse in the one-dimensional viscous Burgers’ equation, which serves as a minimal nonlinear system for testing accuracy against known dynamics. The second configuration considers a two-dimensional laminar, stably stratified channel flow with a localized heat source, providing a more physically relevant scenario in which coupling between momentum and scalar transport plays a central role. Taken together, these examples demonstrate the applicability of the method across systems of increasing complexity.

\subsection{Viscous Burger's equation}
\label{sec:burgers}

We take the 1D viscous Burgers equation
\[
\frac{\partial u}{\partial t} + u\frac{\partial u}{\partial x} \;=\; \nu\,\frac{\partial^2 u}{\partial x^2}, \quad x\in [0,2\pi] = \mathbb{V},\ t>0,
\]
with a baseline solution \(u(x,t)\) satisfying \(u(x,0)=u_0(x)\).
Let the perturbed solution be \(\tilde u(x,t)\) with initial condition that contains the source,
\[
\tilde u(x,0) \;=\; u_0(x) + S(x), \qquad 
S(x)= \frac{I_s}{\sqrt{2\pi}\,\sigma}\exp\!\Big(\displaystyle -\frac{(x-x_s)^2}{2\sigma^2}\Big).
\]
Define the deviation \(q(x,t):=\tilde u(x,t)-u(x,t)\). Subtracting the baseline equation from the perturbed one yields a form,

\begin{equation}
\frac{\partial q}{\partial t} + u\frac{\partial q}{\partial x} + q \frac{\partial u}{\partial x} + q\frac{\partial q}{\partial x} \;=\; \nu\frac{\partial^2 q}{\partial x^2}, \qquad q(x,0)=S(x).
\label{eq:burgers-pert}
\end{equation}
Equivalently, in conservative form, while encoding the initial condition as an instantaneous source at \(t=0\),
\eqref{eq:burgers-pert} is equivalently (in the sense of distributions) written in the form of \eqref{eq:governing_general},
\begin{equation}
\frac{\partial q}{\partial t} + u \frac{\partial q}{\partial x} + q \frac{\partial u}{\partial x}  + q\frac{\partial q}{\partial x} - \nu\frac{\partial^2 q}{\partial x^2} \;=\; S(x)\,\delta(t).
\label{eq:burgers-impulse}
\end{equation}
We pick the initial condition $u_0(x) = 1+\sin(3x), 0 \le x \le 2\pi$, which is a smooth but nontrivial wave profile that is commonly used to probe nonlinear dynamics. In particular, it allows us to assess how the model captures nonlinear wave propagation and the subsequent steepening of gradients. This choice provides a simple and well-controlled setting to evaluate the nonlinear behavior of the proposed method.
The simulations are carried out over $t \in [0,1]$, which is sufficient to capture the relevant nonlinear evolution induced by the initial perturbation while avoiding long-time shock-like behaviors that are not the focus of the present study.
Observations were made at the final time $t=T$ at $x_m =\displaystyle \{0, \frac{2}{5}\pi, \frac{4}{5}\pi , \frac{6}{5}\pi, \frac{8}{5}\pi\}$. I.e., five sensors evenly distributed in the computational domain and ${M}_j = \delta(x-x_{m,j}) \delta(t-T)$. 
The source is placed at $x_s=3$ with the shape of a Gaussian,
$K(x_s) = \displaystyle \exp{\left(\displaystyle-\frac{(x-x_s)^2}{2}\right)}$, while the source intensity is chosen as $I_s = 0.3$.
The adjoint fields were solved from the following adjoint equation
\begin{equation}
    \frac{\partial q_j^{\dagger}}{\partial \tau} - u\frac{\partial q_j^{\dagger}}{\partial x} - \nu \frac{\partial^2 q^{\dagger}_j}{\partial x^2} = 0, \quad q_j^\dagger(x, \tau=0) = {M}_j(x), \quad j= 1,2,\ldots N_m,
\end{equation}
where $\tau$ is the reverse time axis $\tau = T-t$.
The linear positional embedding are evaluated by,
\begin{equation}
    \boldsymbol{s}^{\dagger}(x) = \int_0^T\delta(t) \boldsymbol{q}^{\dagger}dt = \boldsymbol{q}^{\dagger}(x,\tau = T), \quad \tilde{\boldsymbol{s}}^{\dagger}(x_s) = \int_0^{2\pi} K(x_s) \boldsymbol{q}^{\dagger} dx, \;\;\forall x_s \in \mathbb{V}.
\end{equation}
In addition, the eigenmodes $\psi_{j,k}$ of the Hessian are computed using the algorithm presented before.
The viscous Burgers equation is solved using a finite-difference scheme in conservative form, with the diffusive term treated implicitly to enhance numerical stability. The adjoint Burgers equation is discretized using a discretize-then-transpose approach, which guarantees that the forward–adjoint duality relation is satisfied to machine precision.

\begin{figure}
    \includegraphics[width=0.9\linewidth]{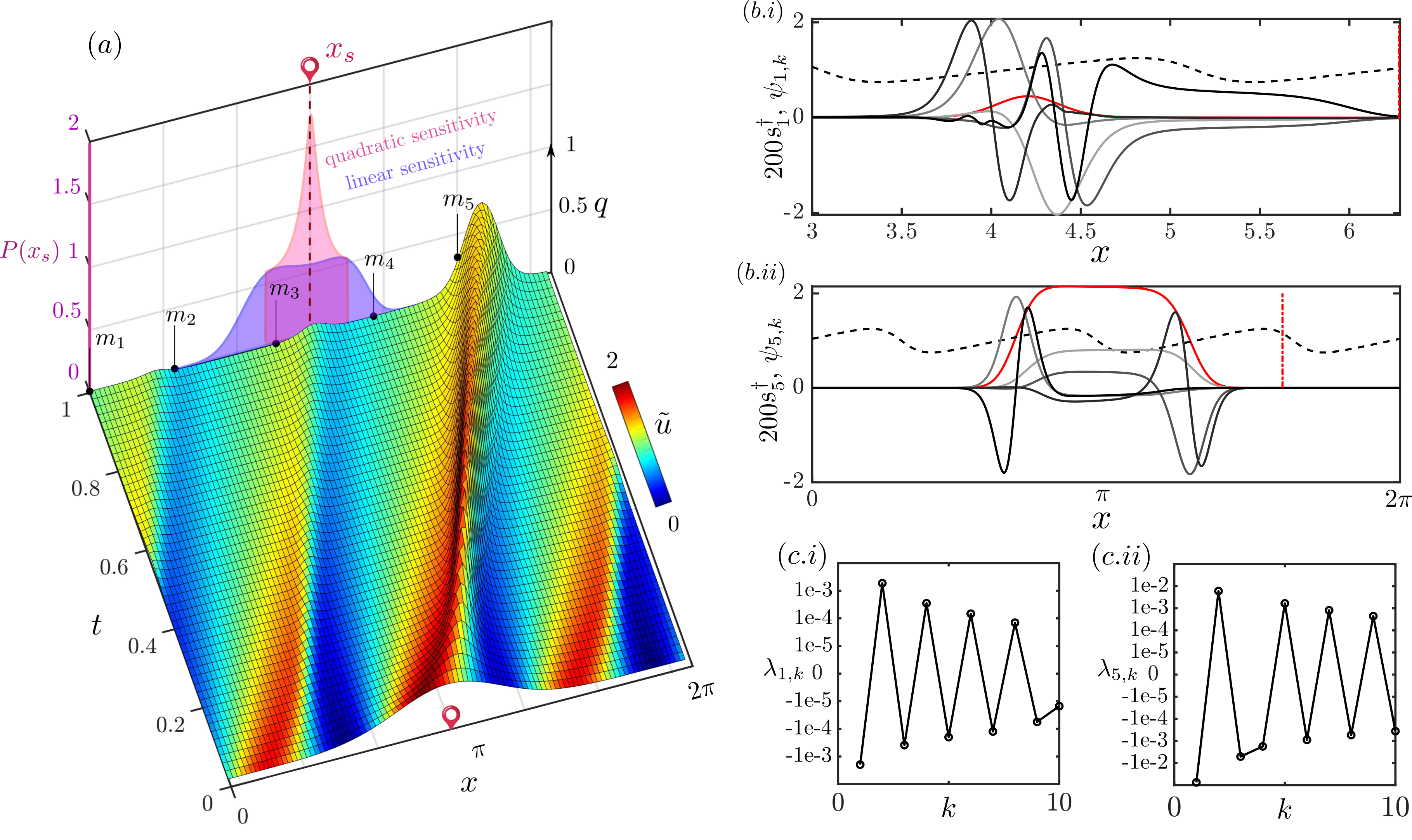}
    \caption{
        (a) The surface plot of $q$, colored by the evolution of the total forward field $\tilde{u}$, with the resulting probability distribution of the source, $P(x_s)$ using linear embedding (equation \ref{eqn:linear_angle}), shown in blue patch, and quadratic embedding (equation \ref{eqn:quad_angle}) in red patch. (b.i) Scaled adjoint field $200 s^{\dagger}_1$ (red)  and five leading eigenmodes of the Hessian, $\psi_{1k}$ (black) from the first sensor. The black dashed lines mark the base state $u$ at the measurement time. (b.ii) Scaled adjoint field $200 s^{\dagger}_5$ (red) and five leading eigenmodes of the Hessian, $\psi_{5k}$ (black) from the last sensor. (c.i) Leading eigenvalues ($\{\lambda_{1k}\}_1^{10}$) of the Hessian for the first sensor. (c.ii) Leading eigenvalues ($\{\lambda_{5k}\}_1^{10}$) of the Hessian for the last sensor.}
    \label{fig:burgers_eigenmodes}
\end{figure}

The forward and perturbed solutions are illustrated in figure \ref{fig:burgers_eigenmodes}$(a)$. The small Gaussian perturbation added at $x_s=3$ modifies both the initial and final states. The perturbation field $q=\tilde{u} - u$ evolves from a Gaussian shape at $t=0$ into a broader, disconnected structure at $T=1$, reflecting both convection by the background flow and viscous diffusion. The five measurement locations, marked in black vector pins, are chosen to sample this distorted perturbation field at distinct phases of its evolution. Panel (a) further compares the linear and quadratic embeddings in terms of the inferred source probability distribution $P(x_s)$, evaluated from equation \eqref{eqn:probability}. The quadratic embedding using $N_{eig} = 5$ produces a sharper peak near the true source location, reducing spurious spreading of the posterior distribution.

The corresponding adjoint field $\boldsymbol{s}^{\dagger}$ and the eigendecomposition of the Hessian are shown in figure \ref{fig:burgers_eigenmodes}. Panels (b.i)–(b.ii) and (c.i)–(c.ii) highlight results from the first and last sensors, respectively. In both cases, the scaled adjoint fields $s_j^\dagger$ reveal the spatial support of each sensor’s dependence, while the leading eigenmodes $\psi_{j,k}$ represent directions in source space that are most strongly amplified in the quadratic response, {$\frac 12 \mathcal{H}[S,S]$} in \eqref{eqn:quadratic_embedding}. The eigenvalue spectra have both positive and negative values, and are rapidly decaying, suggesting that only a handful of modes are required to capture the dominant nonlinear sensitivity.

\begin{figure}[h]
    \includegraphics[width=0.9\linewidth]{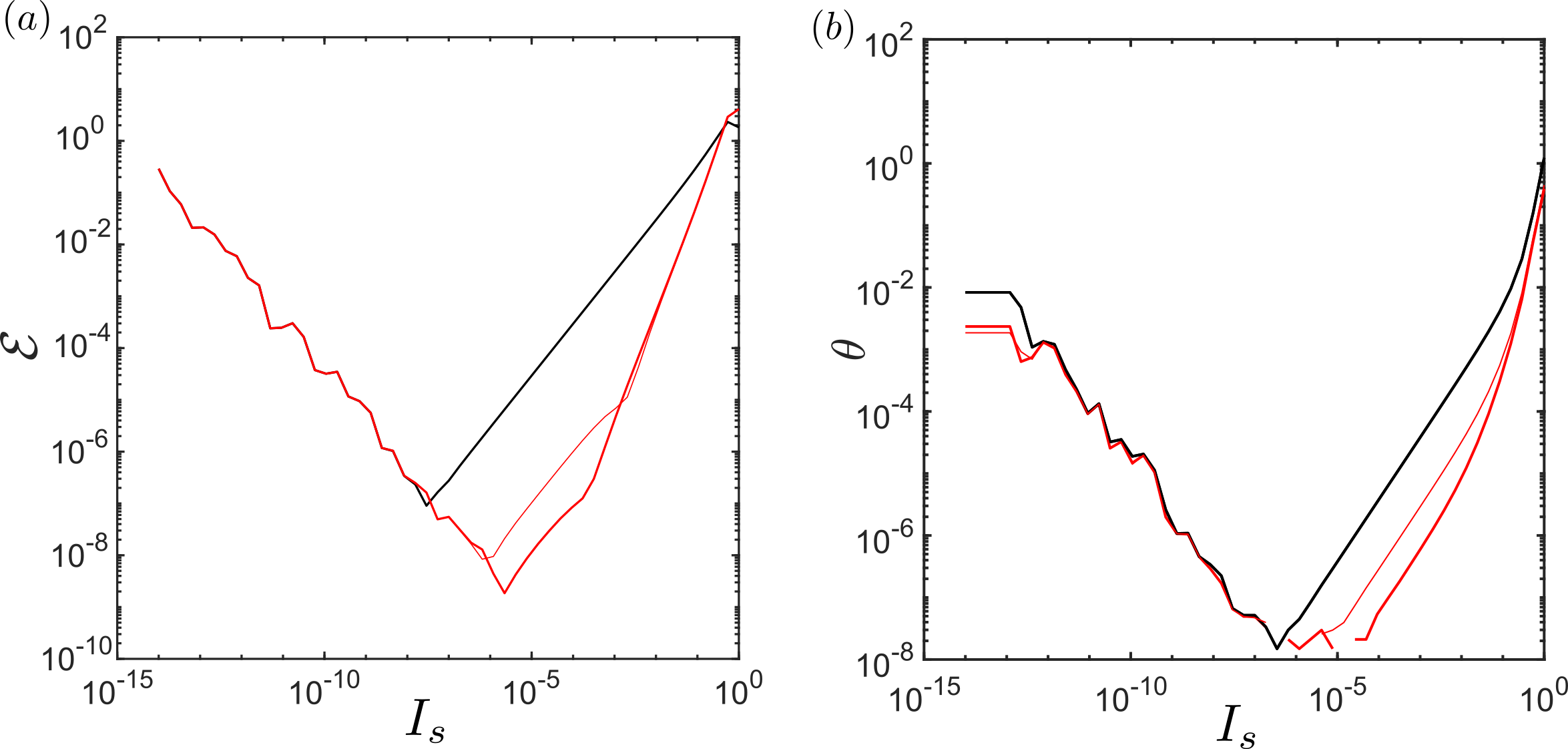}
    \caption{Taylor test for the accuracy of linear and quadratic embeddings, showing (a) the relative difference in the measurement and its approximation using duality relations, and (b) angle between the embedding and the measurement, as a function of the source intensity $I_s$. The linear embedding is shown by the black line, while quadratic embeddings with five and ten eigenmodes are shown by the thin and thick red lines, respectively.}
    \label{fig:burgers_taylor}
\end{figure}
The accuracy of the linear and quadratic embeddings is assessed using a Taylor test (figure \ref{fig:burgers_taylor}), in which the source intensity $I_s$ is systematically reduced and the validity of equation \ref{eq:taylor_expansion} is examined with zero (linear embedding only), five, and ten eigenmodes. 
The error is quantified as,
\begin{equation}
    \mathcal{E} = \frac{||\boldsymbol{m} - I_s\tilde{s}_j^{\dagger}(\boldsymbol{x}_s) - \frac 12 I_s^2 \tilde{h}_j(\boldsymbol{x}_s)||}{||\boldsymbol{m}||},
    \label{eq:taylor_error}
\end{equation}
Panel (a) shows that the quadratic expansion attains markedly smaller relative errors, demonstrating second-order accuracy across a range of source intensities. Panel (b) further indicates that the angle between the embedded approximation and the true measurement remains consistently small.
Both errors rise up for very small $I_s$ due to the round-off error while evaluating the nominator in equation \ref{eq:taylor_error}, which cannot be smaller than machine zero $\mathcal{O}(10^{-15})$.
For the toy problem, retaining only five Hessian eigenmodes already provides a substantial improvement over the linear approximation, with additional modes yielding further refinement.

\begin{figure}
    \includegraphics[width=0.99\linewidth]{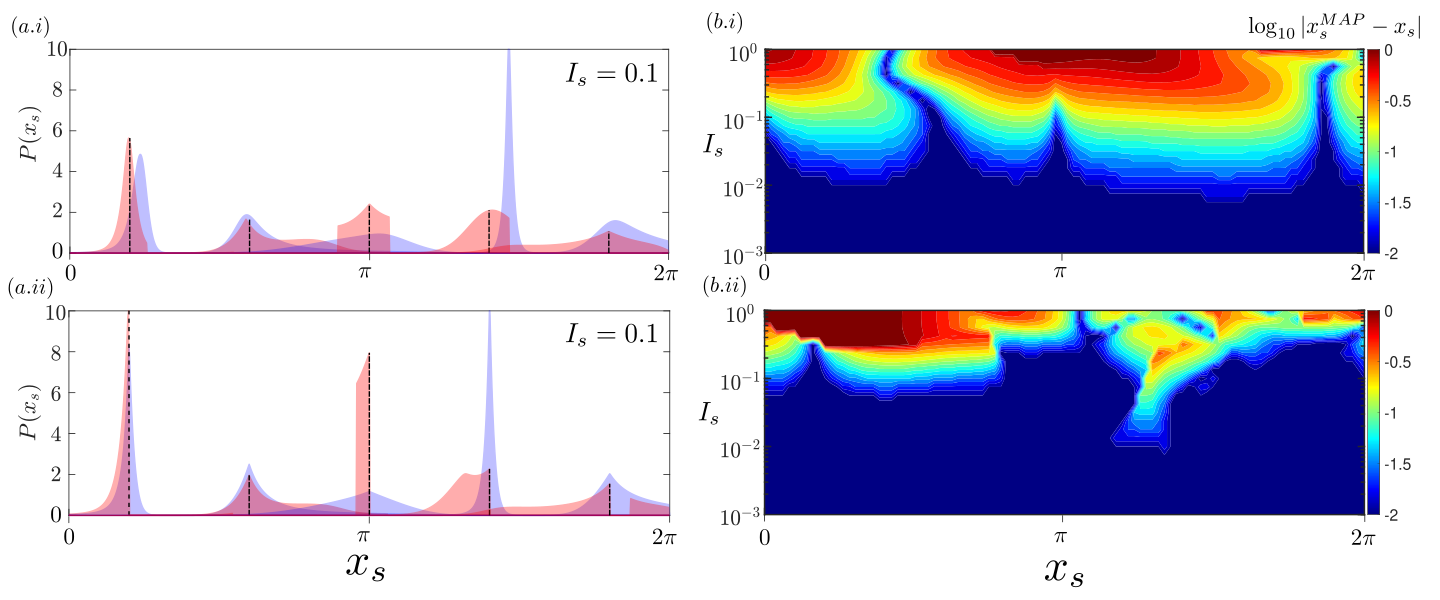}
    \caption{(a.i) Predicted probability distribution $P(x_s)$ for selected source locations (black dashed lines) with moderate intensity $I_s=0.1$. Blue and red shades show results from linear and quadratic embeddings. black dashed lines mark the true source locations. (a.ii) As in (a.i), but for a smaller source intensity $I_s=0.01$. (b.i) Log-MAP distance as a function of source location $x_s$ and source intensity $I_s$ using linear embedding. (b.ii) As in (b.i), but for quadratic embedding with five Hessian eigenmodes.}

    \label{fig:burgers_accuracy}
\end{figure}
Figure \ref{fig:burgers_accuracy} assesses the accuracy of source recovery across a range of test cases. 
The posterior probability of the source location, $P(\boldsymbol{x}_s)$, is shown for the linear positional embedding (blue shades) and the quadratic embedding (red shades). Regions corresponding to negative inferred intensity coefficients are deemed unphysical and are therefore masked according to \ref{eqn:probability}.
At low source intensity ($I_s = 0.01$, panel a.ii), nonlinear effects are weak and the quadratic correction is negligible; both embeddings yield posterior distributions centered at the true source location.
In contrast, at moderate source intensity ($I_s = 0.1$, panel a.i), nonlinear effects become appreciable. In this regime, the quadratic embedding produces sharply localized peaks in $P(\boldsymbol{x}_s)$ at the true source positions, whereas the linear embedding exhibits small but systematic offsets from the exact location. This behavior reflects the finite-amplitude nature of the source within the weakly nonlinear regime, for which the quadratic correction complements the linear positional embedding and more accurately represents the effective embedding space.

The error in the MAP estimation, $|x^{MAP}_s - x_s|$ (shown in panels b.i–b.ii) provides a systematic comparison across $(x_s,I_s)$.
For weak sources ($I_s < 0.01$), both linear and quadratic embeddings correctly predict the source location across the domain. As the source intensity increases to moderate levels ($0.01 < I_s < 0.2$), the quadratic embedding maintains good reconstruction accuracy, while the linear embedding generally begins to degrade, except near specific locations (notably $x \approx 2$ and $6$).
At larger source intensities, the performance of the two embeddings diverges: regions in which the linear embedding successfully predicts the source location often correspond to poor performance of the quadratic embedding, and vice versa. This behavior may be related to the interaction between the base initial condition and the finite observation time window, which together generate spatially non-uniform nonlinear effects.

Overall, these results demonstrate that for the simple one-dimensional viscous Burgers equation, nonlinear source effects can be accurately captured by the quadratic positional embedding, which plays a measurable role in shaping the posterior distribution beyond the linear adjoint framework.

\subsection{Nonlinear source localization in stratified channel flow}
\label{sec:stratifiedchannel}
Having established the methodology in a simplified setting, we now turn to a more realistic fluid system: a stratified half channel flow with a localized heat source. This configuration introduces additional physical complexity through buoyancy coupling and background temperature stratification, providing a more stringent test for the quadratic embedding framework. Under the Boussinesq approximation, the nondimensional governing equations can be written as,
\begin{equation}
\begin{aligned}
    \frac{\partial \mathbf{u}}{\partial t} + \mathbf{u} \cdot \nabla \mathbf{u} + \nabla p - \frac{1}{Re}\nabla^2 \mathbf{u} + Ri\,c\, \mathbf{e}_y &= \boldsymbol{0}, 
    \qquad \nabla \cdot \mathbf{u} = 0, \\
    \frac{\partial c}{\partial t} + \mathbf{u} \cdot \nabla c - \frac{1}{Pe}\nabla^2 c &= I_s \,\delta(\boldsymbol{x}-\boldsymbol{x}_s),
\end{aligned}
\label{eq:channel-boussinesq}
\end{equation}
where $\mathbf{u}(\boldsymbol{x},t)$ is the velocity field, $p(\boldsymbol{x},t)$ is the pressure, and $c(\boldsymbol{x},t)$ is the scalar (temperature fluctuation) field generated by a steady, point source $S=I_s \delta(\boldsymbol{x}-\boldsymbol{x}_s)$ at location $\boldsymbol{x}_s$.  
The nondimensional parameters appearing in \eqref{eq:channel-boussinesq} are the Reynolds number \(Re=UL/\nu\), Peclet number \(Pe=UL/\kappa\), and Richardson number \(Ri=(g \alpha \Delta C L)/U^2\), which quantifies the relative importance of buoyancy to inertia. The reference scales are chosen as follows: the half-channel height \(L\) serves as the characteristic length, the bulk velocity \(U\) of the laminar Poiseuille base flow defines the velocity scale. Pressure is nondimensionalized by \(\rho U^2\), while the scalar field \(c\) is scaled by the background temperature difference \(\Delta C\). The material properties are characterized by the kinematic viscosity \(\nu\) and the scalar diffusivity \(\kappa\), the thermal expansion coefficient of the fluid, $\alpha$, which appear in the definitions of \(Re\), \(Pe\), and \(Ri\), respectively.
These physical parameters are set to $Re=500$, $Pe=500$, and $Ri = 5$. The corresponding Froude number is $Fr=\sqrt{1/Ri}=0.45$, ensuring a strongly stratified configuration.

The simulation is performed with the finite volume method. For spatial discretizations, second-order interpolations are used to obtain flux at the cell boundaries. For time-stepping, the Adams-Bashforth scheme is utilized for the advection terms, and Crank-Nicolson is used for the diffusion terms. Pressure projection and implicit diffusion are treated by inverting the modified wave number in Fourier space to enhance numerical efficiency. The accuracy is second-order in space and time.

In the numerical experiments presented here, the computational domain is a two-dimensional channel of size $L_x=3\pi$ and $L_y=1$, discretized using $N_x=192$ and $N_y=64$ grid points.
The computational domain is periodic in the streamwise (\(x\)) direction, while no-slip boundary conditions are imposed at the lower (\(y=0\)) channel walls. The scalar field \(c\), which represents temperature fluctuations about a stably stratified background, satisfies homogeneous Neumann boundary conditions at the walls, ensuring no flux of heat through the channel boundaries.
The same set of boundary conditions is applied consistently to both the forward and adjoint problems. The flow is integrated with a timestep $\Delta t = 0.004$ up to a final time $t=T$. 
A steady, localized source of intensity \(I_s\) is placed at position \(\boldsymbol{x}_s\), modeled as a Dirac delta forcing term in the scalar transport equation. This forcing perturbs the laminar base state and induces a coupled velocity--scalar response through buoyancy effects captured by the Richardson number.  
The source is represented numerically as a steady Gaussian-smoothed delta distribution centered at $\boldsymbol{x}_s$, namely
\[
    S(\boldsymbol{x}) = 
    I_s \exp\!\left(-\frac{|\boldsymbol{x}-\boldsymbol{x}_s|^2}{2\sigma_s^2}\right), \quad J(t) = \boldsymbol{1}_{t\geq0},
\]
where $\sigma_s$ is the prescribed source width. 
Each sensor kernel is modeled analogously as a Gaussian function of width $\sigma_m$, centered at the measurement locations $\{\boldsymbol{x}_{m,j}\}_{j=1}^{N_m}$, taking the measurement at time $t=T$,
\[
    {M}_j(\boldsymbol{x}) \;=\; 
    \frac{1}{2\pi \sigma_m^2}
    \exp\!\left(-\frac{|\boldsymbol{x}-\boldsymbol{x}_{m,j}|^2}{2\sigma_m^2}\right)\delta(t-T).
\]
Both $\sigma_s$ and $\sigma_m$ are chosen to be $0.1$ throughout this section.
We deliberately refrain from normalizing the source kernel. Normalization would scale the effective source intensity as $\mathcal{O}(\sigma_s^{-2})$. By keeping the Gaussian kernel unnormalized, we focus on probing nonlinear source effects rather than enforcing exact conservation.

For any given measurement kernel $M_j$, the adjoint equations are given by,
\begin{equation}
\begin{aligned}
    \frac{\partial \mathbf{u}_j^\dagger}{\partial (-t)}
    - (\mathbf{u}\cdot\nabla)\mathbf{u}_j^\dagger
    + (\nabla \mathbf{u})^\top \mathbf{u}_j^\dagger
    + (\nabla c_j)\, c_j^\dagger
    &= \nabla p_j^\dagger + \frac{1}{Re}\nabla^2 \mathbf{u}_j^\dagger , \quad \nabla\!\cdot\!\mathbf{u}_j^\dagger = 0,\\
    \frac{\partial c_j^\dagger}{\partial (-t)}
    - \mathbf{u}\cdot\nabla c_j^\dagger
    - \frac{1}{Pe}\nabla^2 c_j^\dagger
     + Ri\,\mathbf{e}_y\!\cdot\!\mathbf{u}_j^\dagger &= {M}_j.
\end{aligned}
\label{eq:adjoint-boussinesq}
\end{equation}
These equations are derived from the forward-adjoint duality relation \eqref{eqn:linear_sensitivity_new} using integration by parts, and are solved backward in time.
In the current paper, we implement the discrete-then-transpose approach, i.e., the ``discrete adjoint" to ensure the accuracy of the adjoint fields \citep{wang2019discrete}.
We can define the adjoint source $s^{\dagger}_j = \int_0^T c^{\dagger}_j d\tau$, following the definition in equation \eqref{eqn:linear_sensitivity_new}.

\begin{figure}
    \centering
    \includegraphics[width=\linewidth]{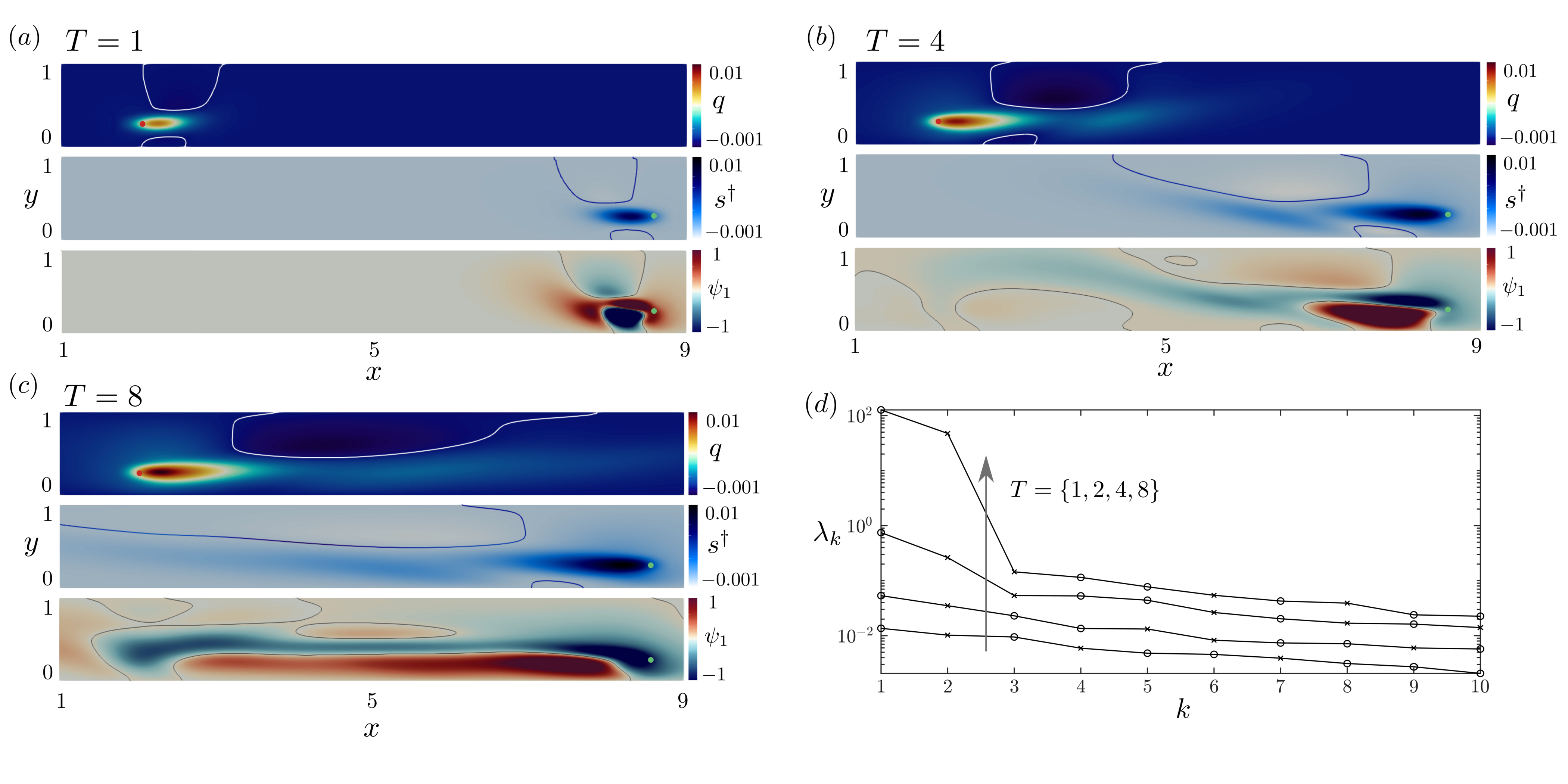}
    \caption{(a-c) Forward temperature field $c(\boldsymbol{x},T)$, first-order sensitivity $s^{\dagger}$, and leading eigenvector of the second-order sensitivity $\psi_1$ for time horizons $T=\{1,4,8\}$. Contours indicate the zero-level curves of the respective fields. (d) The eigenvalues of the Hessian matrix $\mathcal{H}$ for the same sensor but with different time horizons $T=\{1,2,4,8\}$, while we show the data with absolute values, the negative ones are marked with crosses, while the positive ones are marked with circles.
}
    \label{fig:stratified_evolution}
\end{figure}

\subsubsection{Sensor sensitivity analysis}
We first examine the evolution of sensor sensitivity in time.
Figure \ref{fig:stratified_evolution} illustrates the scalar field, adjoint sensitivity field, and the leading eigenmode of the Hessian for measurement times $T=\{1,4,8\}$, and for source and sensor located at the same height $y_s = y_m = 0.33$. The forward scalar field exhibits characteristic lee-wave patterns downstream of the localized heat source, resembling a buoyancy ``blockage" in which stratification strongly modulates the propagation of scalar disturbances. These oscillatory features are a direct consequence of the stable background stratification and provide a persistent signature of the source even at long times.

The adjoint sensitivity fields highlight the regions where sensors placed downstream are most responsive to perturbations upstream. At later times, the sensitivity shifts along wave crests aligned with a reverse lee-wave pattern. This behavior reflects the physical pathway by which scalar perturbations are transported toward the sensors in stratified flows. It is interesting that the forward and adjoint scalar fields both move away from the wall, unlike the unstratified counterpart \citep{wang_hasegawa_zaki_2019}.

The leading Hessian eigenmode reveals a qualitatively distinct structure. Unlike the first-order adjoint sensitivity, which vanishes in regions orthogonal to the linear influence of the source, the Hessian mode can maintain finite amplitude where the first-order sensitivity vanishes, which we will see in more detail in later sections. This phenomenon indicates that the quadratic embedding captures second-order source–field interactions that are invisible to the linear adjoint framework. Physically, this corresponds to a broader footprint of source detectability: regions that appear insensitive at first order may still contribute to the sensor signals through nonlinear effects, which can be captured by the quadratic embedding. Consequently, the quadratic embedding extends the effective domain of sensitivity and enables sharper and more robust source localization.
From the eigenvalue distribution of the Hessian $\mathcal{H}$, we observe that the second-order sensitivity grows exponentially as the time horizon increases, $T=\{1,2,4,8\}$. This behavior was previously reported in turbulent flows, where the Hessian eigenvalues grow with a rate equal to twice the Lyapunov exponent \citep{wang_wang_zaki_2022}. Here, we find the same phenomenon in laminar flows, indicating that nonlinear effects amplify exponentially with time. This provides further motivation for employing quadratic embeddings to quantify such effects.

\begin{figure}[h]
    \centering
    \includegraphics[width=\linewidth]{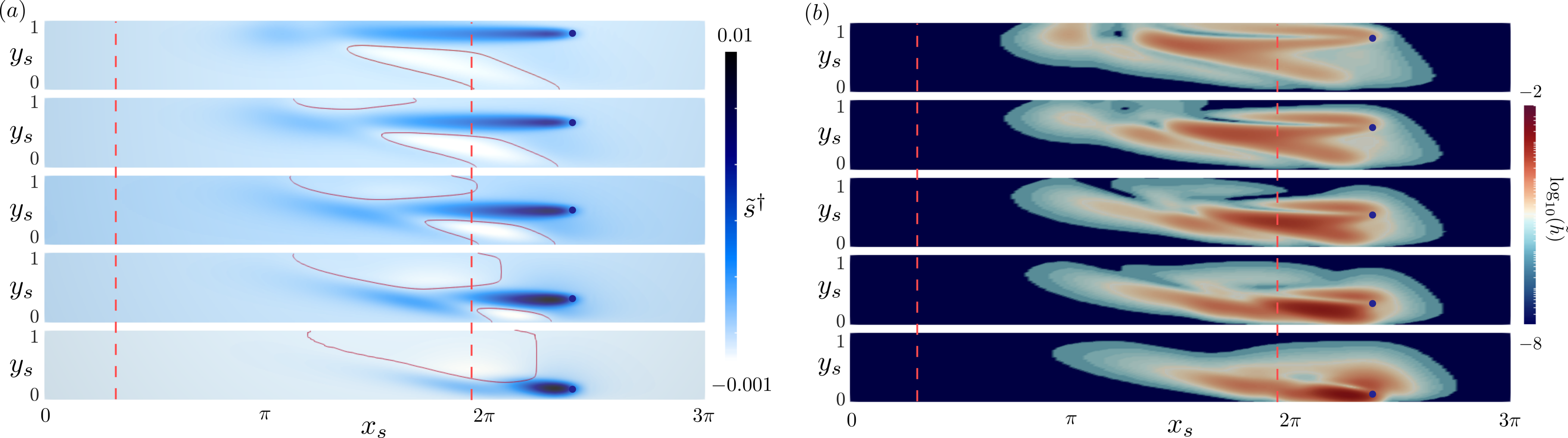}
    \caption{(a) The linear embedding fields $\tilde{s}^{\dagger}_k$ from selected sensors located as $x_m = 7.5$. Dark dots mark the location of the sensors, while the red solid line marks the region of zero linear sensitivity. The dashed line marks the boundary of the search region. (b) The quadratic embedding $\tilde{h}^{\dagger}_k$ from the same set of selected sensors, shown on a log scale.}
    \label{fig:stratified_sensitivity_sensorarray}
\end{figure}
We focus on time $T=4$, where the sensitivity fields extend sufficiently upstream while remaining unaffected by the periodic boundary condition. To examine the effect of sensor at different heights, we consider sensors positioned at different vertical locations along $x_m = 7.5$ and plot the linear sensitivity embedding $\tilde{s}$ and $\tilde{h}$ for these sensors. The results are shown in figure \ref{fig:stratified_sensitivity_sensorarray}.
This sensor configuration yields nearly vanishing linear sensitivity to upstream perturbations located at $x_s = 1.5$ (a distance $x_m - x_s = 6$). The corresponding zero-sensitivity lines are highlighted in panel (a). Upstream of the sensors, these lines incline relative to the horizontal, creating a common region where all sensors exhibit very limited linear sensitivity.
In contrast, the quadratic embeddings remain positive within these regions. This observation highlights the crucial role of quadratic embeddings in recovering source information when linear sensitivity alone fails.
These quadratic embeddings are evaluated using five leading eigenmodes of the Hessian. Increasing that number to ten yields nearly identical results.

\begin{figure}[h]
    \centering
    \includegraphics[width=\linewidth]{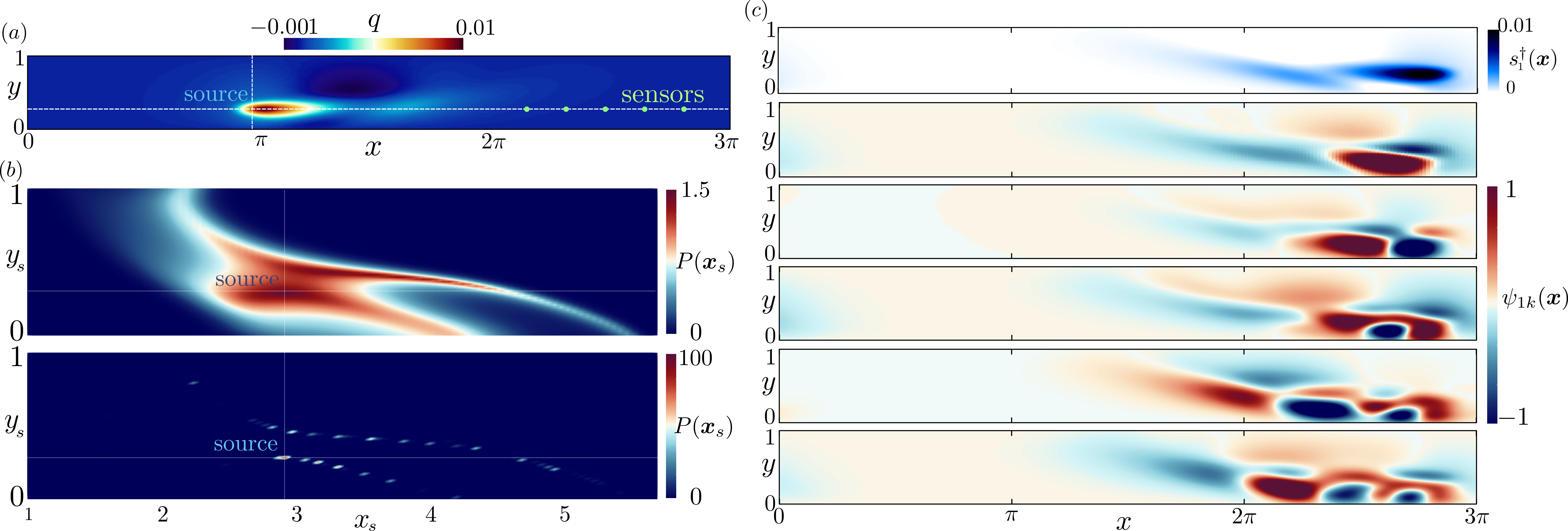}
    \caption{(a) Forward field $c$ at $T=4$ for a heat source in a temperature-stratified laminar channel flow, with five downstream sensors marked by green dots.  
(b) Reconstructed probability distribution of the source location $P(x_s)$ using linear (top) and quadratic (bottom) embeddings. The quadratic embedding produces a distribution that is markedly more concentrated around the true source.  
(c) Adjoint field from the farthest sensor, $s^{\dagger}_1$, together with the five leading eigenmodes of the Hessian associated with the same sensor.
}
    \label{fig:Stratified_sample}
\end{figure}

\subsubsection{Source inference results}
\label{sec:channel_reconstruction}
We consider a sample case where the intensity of the source is $I_s = 0.05$, the measurement time $T=4$, and a 5-sensor array is arranged at the same height as the source ($y_s = y_m = 0.33$), as shown in figure \ref{fig:Stratified_sample}.
The forward field and source reconstruction results for this setup are summarized in figure \ref{fig:Stratified_sample}. Panel $(a)$ shows the normalized perturbation field $c$ at $T=4$, where the injected heat source generates a plume that is advected downstream while being diffused by viscosity and stratification. The reconstructed source distributions in panel $(b)$ highlight the improvement achieved by the quadratic embedding: while the linear embedding produces a relatively broad source probability distribution, the quadratic correction sharpens the distribution and yields a peak concentrated near the true source. Panel $(c)$ displays the adjoint field from the farthest sensor, $s^{\dagger}_1$, along with the five leading eigenmodes of the associated Hessian. 
Results for other sensors are just circular translations of the first sensor due to the translation-equivariant nature of the setup.
The adjoint field highlights the extended upstream sensitivity of the measurement, while the Hessian eigenmodes identify dominant directions in source space that govern the nonlinear response. Higher-order modes are increasingly oscillatory, providing finer resolution for distinguishing nearby sources. All eigenmodes are inclined with respect to the horizontal, a feature that shapes the reconstructed source probability distribution. Notably, both linear and quadratic methods indicate that sources aligned with the backward lee-wave direction are the most difficult to resolve, while incorporating the quadratic information substantially enhances source localization in stratified flows.

\begin{figure}
    \centering
    \includegraphics[width=0.9\linewidth]{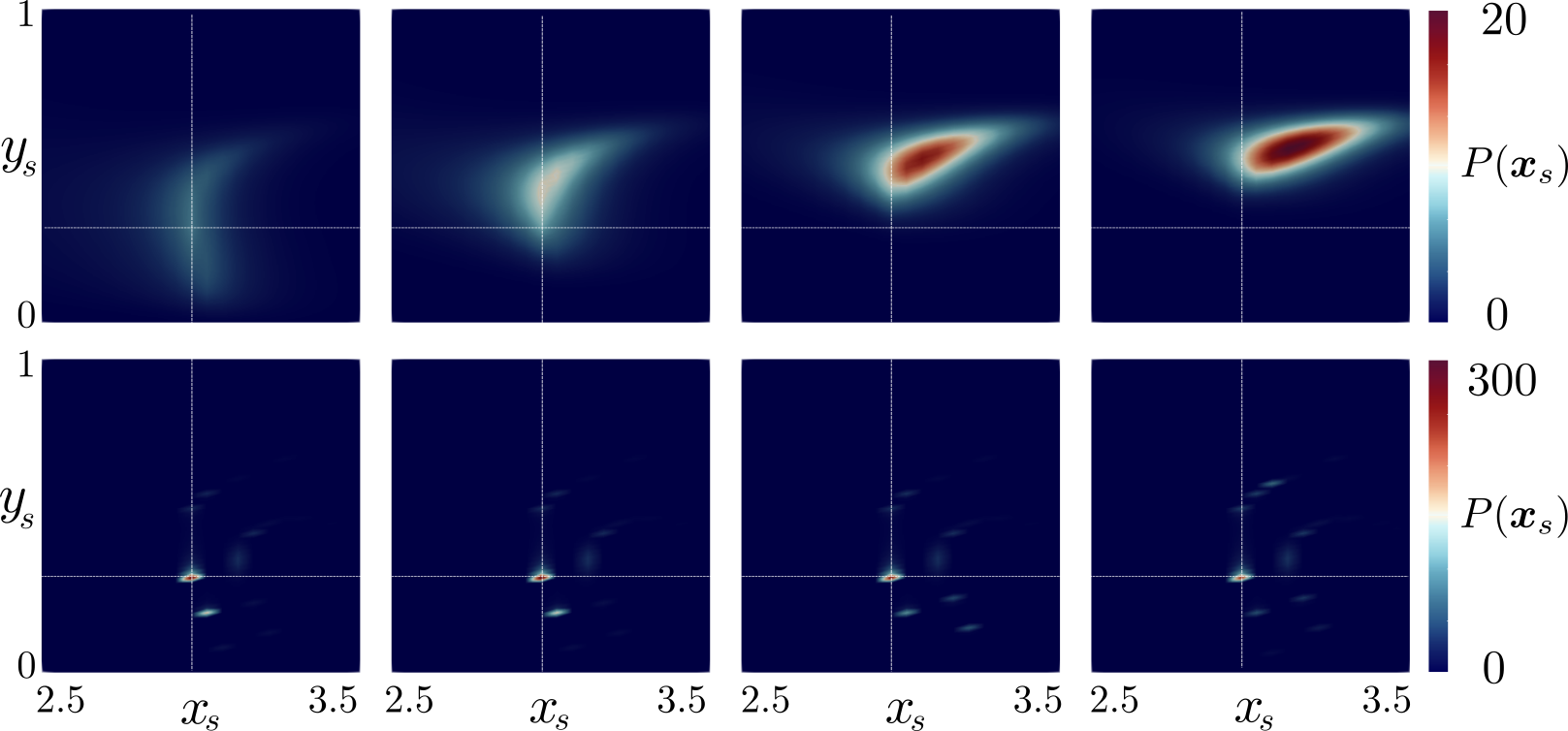}
    \caption{Probability distribution of reconstructed source location using linear (top) embedding and quadratic (bottom) embeddings. From left to right the intensity of the source increases, $I_s = \{0.002, 0.02, 0.1, 0.2\}$.}
    \label{fig:stratified_nonlinearity}
\end{figure}

In order to examine the power of sensing a whole vertical distribution of scalar, we use another configuration with $N_m=32$ sensors evenly distributed along the line $x_m=7.5$, representing the full spatial extent of a remote sensor network at a given streamwise location.
We first pick a source location where both linear and quadratic embedding are finite, namely $\boldsymbol{x}_s = [3, 0.33]^{\top}$.
Figure~\ref{fig:stratified_nonlinearity} shows the reconstructed probability distribution $P(\boldsymbol{x}_s)$ for this source, compared with the ground truth location marked with dashed lines. 
The comparison is made across increasing source intensities, using the linear embedding (top row) and the quadratic embedding (bottom row). For all source intensities, the quadratic embedding remains sharply localized and accurately recovers the true source position. By contrast, the linear embedding becomes increasingly biased as nonlinear effects grow, producing broader posterior distributions that are also displaced away from the ground truth.

\begin{figure}
    \centering
    \begin{minipage}{0.95\linewidth}
        \centering
        \includegraphics[width=\linewidth]{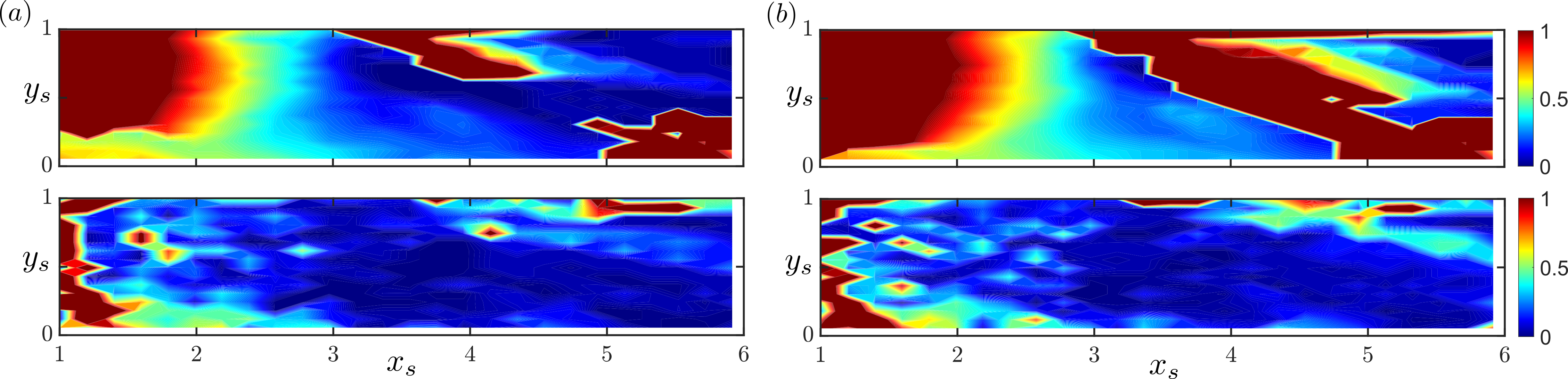}
    \end{minipage}%
    \hfill
    \begin{minipage}{0.04\linewidth}
        \centering
        \rotatebox{270}{\small $||\boldsymbol{x}_s^{MAP} - \boldsymbol{x}_s||$}
    \end{minipage}
    \caption{Error in the location prediction using linear (top) and quadratic (bottom) position embeddings. (a) and (b) show results with different levels of source intensity, $I_s=\{0.1,0.2\}$, respectively.}
    \label{fig:stratified_accuracy}
\end{figure}

We further plot the accuracy of source reconstruction under different levels of nonlinearity, i.e., source intensity $I_s$.
The resulting PDF of reconstructed source is shown in figure \ref{fig:stratified_accuracy}, which plots the error in source localization, $||\boldsymbol{x}_s^{MAP} - \boldsymbol{x}_s||$, for moderate and strong sources ($I_s=0.1,\,0.2$). In both cases, the quadratic embedding achieves significantly higher accuracy than the linear embedding. The accuracy is low when the sources are very far away, indicating decaying sensitivity both for the linear and quadratic embeddings.
Interestingly, the results with linear embeddings are particularly problematic near regions of vanishing sensitivity, as shown in figure \ref{fig:stratified_sensitivity_sensorarray}$(a)$, while the quadratic embeddings yield a much better quality of source inference at those locations.
The quadratic embedding is also able to improve the source localization further away from the sensor array, compared with linear embedding method.

\section{Generalization to multiple sensors}
\label{sec:discussion}
The formulation can be generalized to multiple sources $\{\boldsymbol{x}_s^{(i)}\}_1^{N_s}$ with possibily different intensities $\{I_s^{(i)}\}_1^{N_s}$.
The measurement $m_j$ can be approximated by the positional embedding as,
\begin{equation}
    m_j = \sum_{l=1}^{N_s} I^{(l)}_s \tilde{s}_j^{\dagger} + \frac 12 \sum_{l=1}^{N_s}\sum_{p=1}^{N_s} I_s^{(l)}I_s^{(p)} \underbrace{\sum_{k=1}^N\lambda_k \tilde{\psi}_{j,k}(\boldsymbol{x}_s^{(l)})\tilde{\psi}_{j,k}(\boldsymbol{x}_s^{(p)})}_{\tilde{H}_{lp}}.
\end{equation}
There are $N_s (N_s+1) /2$ vectors within the embedding, namely $ B(\{\boldsymbol{x}^{(l)}_s\}) = \{\tilde{s}_j^{\dagger}, \tilde{H}_{lp}\}_{l,p=1}^{N_s}$ forming a hyperplane that the vector $\boldsymbol{m}$ lives on. 
Therefore, the dimension of $\boldsymbol{m}$, the sensor number $N_m$, has to be more than $N_s (N_s+1) /2$.
We again quantify the angle between the measurement data and the hyperplane as follows,
\begin{equation}
    \theta = \arccos\!\left( 
    \frac{\|\mathcal{P} \boldsymbol{m}\|}{\|\boldsymbol{m}\|}
    \right),
\end{equation}
where $\mathcal{P}$ denotes the orthogonal projection operator onto the hyperplane spanned by the set of linear and quadratic embedding vectors, $B$.
A small value of $\theta$ indicates that the measured data $\boldsymbol{m}$ is well explained by a superposition of $N_s$ sources and their quadratic interactions, whereas a large value suggests inconsistency with the assumed source model.  

The probabilistic interpretation follows the single-source case: we define a likelihood function based on the angular misfit,  
\begin{equation}
    P(\{\boldsymbol{x}_s^{(l)}\}) 
    \;\propto\; 
    \exp\!\left(-\gamma \,\theta(\boldsymbol{m},B) \right),
\end{equation}
where $\theta(\boldsymbol{m},B)$ denotes the angular misfit between the measurement $\boldsymbol{m}$ and the embedding $B$, and the hyperparameter $\gamma$ is chosen as before.  
We further place an exponential prior on the source intensities, equivalently on the projected coefficients $\{\tilde{s}_j\}_{j=1}^{N_s}$.  
This construction naturally penalizes candidate source configurations that lead to incompatible linear–quadratic embeddings.

Geometrically, the quadratic terms $\tilde{H}(\boldsymbol{x}_s^{(l)},\boldsymbol{x}_s^{(p)})$ generate pairwise interaction directions in measurement space, so that the feasible manifold for $\boldsymbol{m}$ grows rapidly in dimension as $N_s$ increases.  
Consequently, the multiple-source inference problem is constrained not only by the number of measurements available but also by the identifiability of individual and pairwise contributions.  
In practice, successful reconstruction requires that the measurement dimension $N_m$ satisfies $\displaystyle N_m \gg \frac{N_s(N_s+1)}{2}$,
ensuring that the linear independence of the embedding vectors is not lost due to projection into a low-dimensional observation space.
\begin{figure}
    \centering
    \includegraphics[width=0.9\linewidth]{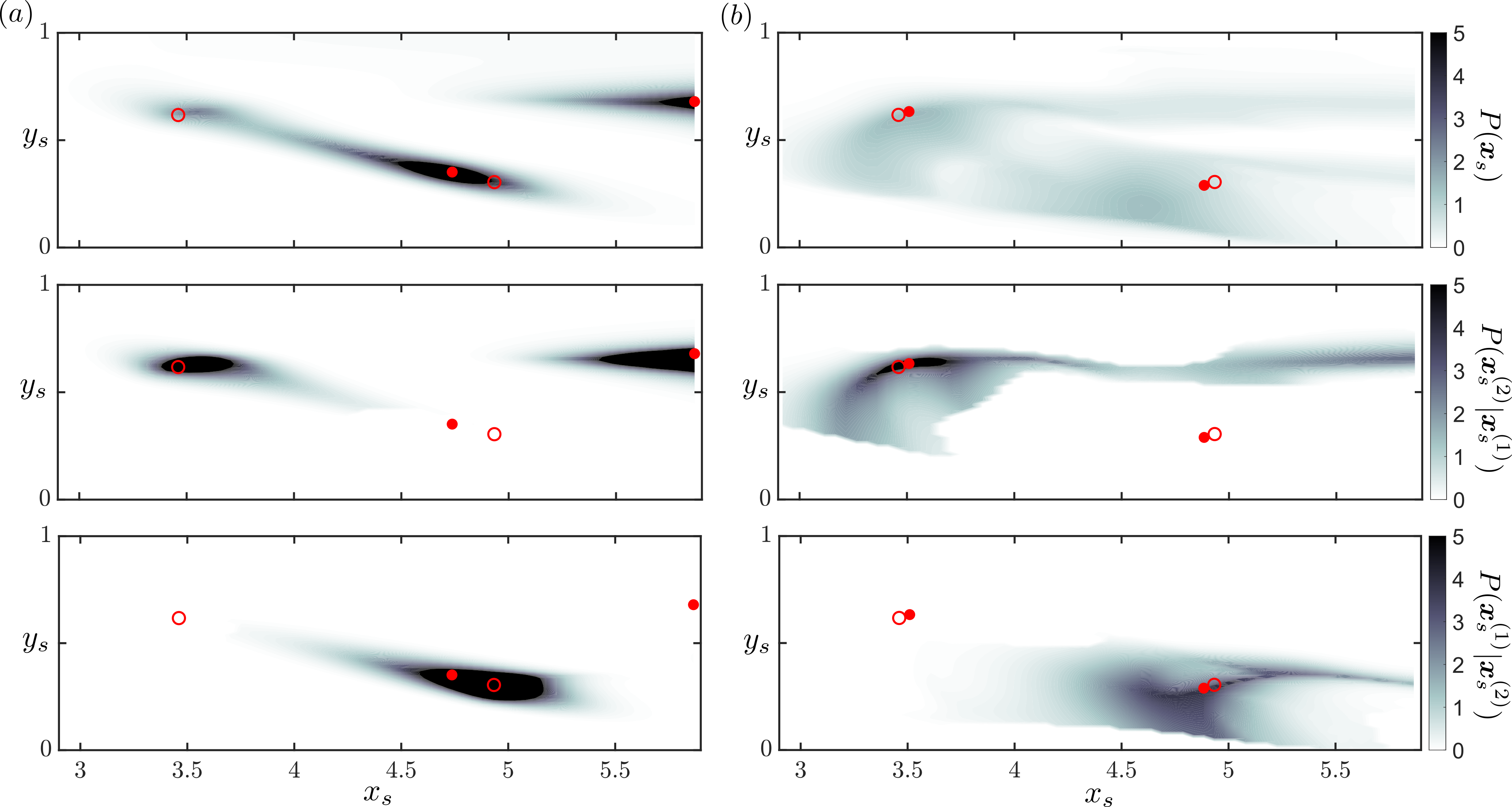}
    \caption{Reconstruction of two sources using (a) linear positional embedding and (b) quadratic positional embedding. From top to bottom: marginal probability $P(\boldsymbol{x}_s)$, conditional probability $P(\boldsymbol{x}_s^{(2)} \mid \boldsymbol{x}_s^{(1)})$, and conditional probability $P(\boldsymbol{x}_s^{(1)} \mid \boldsymbol{x}_s^{(2)})$. Circles denote the true source locations, and filled dots indicate the MAP estimates.}
    \label{fig:stratified_multisource}
\end{figure}

We consider the 32-sensor array at $x_m = 7.5$ with time horizon $T=4$, and introduce two sources in the stratified channel flow: $\boldsymbol{x}_s^{(1)} = [5,\,0.3]^{\top}$ with intensity $I_s^{(1)} = 0.1$, and $\boldsymbol{x}_s^{(2)} = [3.5,\,0.6]^{\top}$ with intensity $I_s^{(2)} = 0.2$. For each sensor, the quadratic embedding is constructed using the five leading eigenmodes ($N_{\mathrm{eig}} = 5$).  

Figure~\ref{fig:stratified_multisource} compares source reconstructions obtained with linear and quadratic embeddings. The top row shows the marginalized probability of one source at a given location,  
\begin{equation}
P(\boldsymbol{x}_s) = \int_{\Omega} P(\boldsymbol{x}_s^{(1)}, \boldsymbol{x}_s^{(2)}) \, d\boldsymbol{x}_s^{(2)} ,
\end{equation}
while the middle and bottom rows show the conditional probabilities given the ground-truth locations of $\boldsymbol{x}_s^{(1)}$ and $\boldsymbol{x}_s^{(2)}$, respectively.  

With linear embedding (panel a), the reconstruction is biased toward spurious regions near the sensor array, yielding MAP estimates that deviate from the true source positions. In contrast, the quadratic embedding (panel b) accurately captures the nonlinear interactions between sources, producing sharply localized posteriors and recovering both positions with high fidelity. These results highlight the essential role of quadratic corrections in resolving multi-source configurations in interactive source environments such as stratified channel flow.

\section{Conclusions}
\label{sec:conclusion}
We have introduced a framework for source detection in nonlinear dynamical systems based on linear and quadratic sensitivity analysis. Extending the classical adjoint formulation, our approach constructs linear positional embeddings from adjoint fields interpreted as Riesz representers, and augments them with quadratic corrections defined by a symmetric bilinear operator and approximated through truncated eigen-expansions. This embedding framework allows measurement data to be projected onto a higher-dimensional subspace that simultaneously accounts for first-order sensitivities and weakly nonlinear interactions, with source inference formulated through principal-angle minimization that admits a natural probabilistic interpretation.  

Applications to benchmark inverse problems, including the viscous Burgers equation and stratified channel flow, illustrate the benefits of this quadratic framework. Linear embeddings provide a useful baseline but fail in regimes where first-order sensitivities vanish or nonlinear effects become pronounced. Incorporating quadratic embeddings resolves these difficulties, yielding sharply localized posteriors and accurate maximum a posteriori estimates. In particular, the channel flow case demonstrates that quadratic terms are essential for disambiguating multiple-source configurations in anisotropic, turbulent environments.  

The methodology operates in a one-shot fashion without iterative refinement of candidate source positions, making it attractive for large-scale or real-time applications. Future work will include a detailed mathematical analysis of measurement uncertainty and probabilistic modeling choices, development of adaptive sensor placement strategies to enhance identifiability, and integration with reduced-order modeling and data assimilation frameworks to enable efficient deployment in complex flow scenarios.
Although the proposed framework naturally admits extensions through higher-order expansions of the nonlinear measurement functional, the computational cost and numerical stability of evaluating higher-order derivatives rapidly become prohibitive, particularly for long assimilation horizons. For this reason, we restrict the present study to a quadratic expansion as a well-posed and robust foundation.
Future research will focus on a rigorous Bayesian analysis of measurement uncertainty and prior modeling choices, and the development of adaptive sensor placement strategies to enhance identifiability.

\begin{acknowledgments}
This work is supported by the National Science Foundation, with Grant No. 2332057 and 2341393.
\end{acknowledgments}

\bibliography{apssamp}

\end{document}